\documentclass[useAMS,usenatbib]{mn2e}
\usepackage{graphicx}
\usepackage{supertabular}

\newcommand{\T}      {T}
\newcommand{\E}      {E}
\newcommand{\BOLTZ}  {\it k}
\newcommand{\C}      {c}
\newcommand{\h}      {h}
\newcommand{\etal}    {{\it et al}}
\newcommand{\abin}    {{\it ab initio}}
\newcommand{\F}      {\nu}
\newcommand{\htdp}    {H$_2$D$^+$}
\newcommand{\dthp}    {D$_2$H$^+$}
\newcommand{\htp}     {H$_3^+$}
\newcommand{\EAC}     {A}
\newcommand{\EBC}     {B}
\newcommand{\wn}      {cm$^{-1}$}
\newcommand{\rotn}    [3]{$#1_{#2#3}$}
\newcommand{\pf}      {z}
\newcommand{\cf}      {W}
\newcommand{\nsdf}    {g}

\title[A computed line list for the \htdp\ molecular ion]{A computed line list for the \htdp\ molecular ion}
\author[T. Sochi, J. Tennyson]{Taha Sochi\thanks{E-mail:
t.sochi@ucl.ac.uk} and Jonathan Tennyson\thanks{E-mail:
j.tennyson@ucl.ac.uk}\\
Department of Physics and Astronomy, University College London, Gower Street, London WC1E 6BT, UK}

\begin{document}

\date{Accepted ---.  Received ---; in original form ---}

\pagerange{\pageref{firstpage}--\pageref{lastpage}} \pubyear{2010}

\maketitle

\label{firstpage}

\begin{abstract}
A comprehensive, calculated line list of frequencies and transition probabilities for the singly
deuterated isotopologue of H$_3^+$, \htdp, is presented. The line list, called ST1, contains over
22 million rotational-vibrational transitions occurring between more than 33 thousand energy
levels; it covers frequencies up to 18500 \wn. All energy levels with rotational quantum number,
$J$, up to 20 are considered, making the line list useful for temperatures up to at least 3000~K.
About 15\% of these levels are fully assigned with approximate rotational and vibrational quantum
numbers. The list is calculated using a previously proposed, high accuracy, \abin\ model and
consistency checks are carried out to test and validate the results. These checks confirm the
accuracy of the list. A temperature-dependent partition function, valid over a more extended
temperature range than those previously published, and cooling function are presented.
Temperature-dependent synthetic spectra in the frequency range $0-10000$~\wn are also given.
\end{abstract}


\section{Introduction}

\htdp\ is one of the simplest polyatomic quantum systems. It consists of two electrons bound to
three nuclei (two protons and one deuteron) forming a triangular shape at equilibrium. The molecule
is an asymmetric prolate top with three vibrational modes: breathing ($\nu_1$), bending ($\nu_2$)
and asymmetric stretch ($\nu_3$). All these modes are infrared active.  \htdp\ also possesses a
permanent electric dipole moment of about 0.6~D due to the displacement of the center of charge
from the center of mass. Hence pure rotational transitions, which occur in the far infrared, can
occur. This all contrasts strongly with the non-deuterated \htp\ molecular ion which has no allowed
rotational spectrum and only one infrared active vibrational mode.

In astrophysical environments, \htdp\ is formed by several reactions; the main one is
\begin{equation}\label{htdpEq}
\textrm{H}_3^+ + \textrm{HD} \rightarrow \textrm{H}_2\textrm{D}^+ + \textrm{H}_2 \, .
\end{equation}
As this and other similar formation reactions are exothermic, the formation of \htdp\ in the cold
interstellar environment is favored. Consequently, the abundance of this species can be greatly
enhanced compared to the underlying D/H ratio. As an isotopologue of \htp, \htdp\ is a major
participant in chemical reactions taking place in interstellar medium. In particular, it plays a
key role in the deuteration processes in these environments \citep{Millar2003}. Because \htp\ lacks
a permanent dipole moment and hence cannot be detected via radio astronomy, its asymmetrical
isotopomers, namely \htdp\ and \dthp, are the most promising tracers of the extremely cold and
highly dense interstellar clouds. These species are the last to remain in gaseous state under these
extreme conditions with observable pure rotational spectra \citep{DalgarnoHNK1973,HerbstM2008Book}.
Of these two isotopomers, \htdp\ is the more abundant and easier to observe and hence it is the
main species to utilize in such astrophysical investigations.

Although the existence of \htdp\ in the interstellar medium (ISM) and astrophysical objects was
contemplated decades ago \citep{DalgarnoHNK1973, DishoeckPKB1992} with some reported tentative
sighting \citep{PhillipsBKWC1985, PaganiWFKGe1992, BoreikoB1993}, it is only relatively recently
that the molecule was firmly detected in the ISM via one of its rotational transition lines
\citep{StarkTD1999}. There have been many subsequent astronomical studies of  \htdp\ spectra
\citep{CaselliTCB2003, VastelPY2004,CeccarelliDLCC2004, StarkSBHDe2004, HogerheijdeCETAe2006,
HarjuHLJMe2006, CernicharoPG2007, CaselliVCTCB2008}. In particular these spectra have been used to
investigate the mid-plane of proto-planetary disks \citep{RamosCE2007}, the kinematics of the
centers of pre-stellar cores \citep{TakCC2005}, and the suggestion of possible use as a probe for
the presence of the hypothetical cloudlets forming the baryonic dark matter
\citep{CeccarelliD2006}.  Vibrational transitions of \htdp\ have yet to be observed astronomically.

The potential importance of \htdp\ spectroscopy for cosmology is obvious as it is a primordial
molecular species which could have a considerable abundance in the early universe. In particular it
has a potential role in the cooling of primordial proto-objects \citep{Dubrovich1993,
DubrovichL1995, DubrovichP2000, GalliP1998,Schleicher08}. Its significance is highlighted by a
number of studies which consider the role of \htdp\ in spectral distortions of Cosmic Microwave
Background Radiation and whether this can be used to determine the \htdp\ and deuterium abundances
at different epochs \citep{DubrovichL1995}. Finally the \htdp\ 372 GHz line has been considered as
a probe for the presence of dark matter \citep{CeccarelliD2006}.

Due to its fundamental and astrophysical importance, \htdp\ has been the subject of a substantial
number of spectroscopic studies. The first successful spectroscopic investigation of \htdp\ was
carried out by \citet{ShyFW1981} where nine rotational-vibrational transitions were measured in the
infrared region between 1800 and 2000 \wn\ using Doppler-tuned fast-ion laser technique, but no
specific spectroscopic assignments were made. This was followed by other spectroscopic
investigations which include the observation and identification of the strong and highly-important
rotational $1_{10}-1_{11}$ transition line of ortho-\htdp\ at 372 GHz by \citet{BogeyDDDL1984} and
\citet{WarnerCPW1984}. \citet{AmanoH2005} measured the frequency of the $1_{10}-1_{11}$ line of
\htdp\ with improved accuracy, as well as the $3_{21}-3_{22}$ line at 646.430 GHz. The three
fundamental vibrational bands of the \htdp\ ion were observed by \citet{AmanoW1984},
\citet{Amano1985} and \citet{FosterMPWPe1986} using laser spectroscopy. \citet{FarnikDKPTN2002}
measured transitions to overtones 2$\nu_2$ and 2$\nu_3$ and to combination $\nu_2+\nu_3$ in
jet-cooled \htdp\ ions. \citet{HlavenkaKPVKG2006} measured second overtone transition frequencies
of \htdp\ using cavity ringdown absorption spectroscopy. Other spectroscopic investigations include
those of \citet{AsvanyHMKSTS2007}, who detected 20 lines using laser induced reaction techniques,
and the recent work of \citet{YonezuMMTA2009}, who reported precise measurement of the transition
frequency of the \htdp\ $2_{12}-1_{11}$ line at 2.363 THz, alongside three more far-infrared lines
of \htdp, using tunable far-infrared spectrometry. However, none of these studies measured absolute
line intensities, although the work of \citet{FarnikDKPTN2002} and \citet{AsvanyHMKSTS2007} give
relative intensities.

Several synthetic, \htdp\ line lists have been generated from \abin\ calculations. Prominent
examples include the line list of \citet{MillerTS1989} and another one generated as part of the
work reported by \citet{AsvanyHMKSTS2007}. Miller {\it et al}'s list extends to rotational level
$J=30$ and covers all levels up to 5500\wn\ above the ground state. These lists were used in a
number of studies for various purposes such as spectroscopic assignment of energy levels and
transition lines from astronomical observations, and for computation of a low-temperature partition
function \citep{SidhuMT1992}. The line lists also played an important role in motivating and
steering the experimental and observational work in this field \citep{MillerTS1989,
AsvanyHMKSTS2007}. Nonetheless, the previous \htdp\ line lists suffered from limitations that
include low energy cut-off and the inclusion of a limited number of levels especially at high $J$.
These limitations, alongside the development of high accuracy \abin\ models of \htp\ and its
isotopologues, including \htdp, by \citet{PolyanskyT1999}, provided the motivation to generate a
more comprehensive and accurate line list. The intention is that the new list will both fill the
previous gaps and provide data of better quality.

The line list, which we call ST1, consists of 22164810 transition lines occurring between 33330
rotational-vibrational levels. These are all the energy levels with rotational quantum number
$J\leq 20$ and frequencies below 18500~\wn.  This line list can be seen as a companion
to the \htp\ line list of \citet{NealeMT1996} which has been extensively used for astrophysical
studies; although for reasons explained below, the ST1 line list is actually expected to be more
accurate.

\section{Method}

Vibration-rotation calculations were performed using the DVR3D code of \citet{TennysonKBHPRZ2004}.
The code calculates, among other things, wavefunctions, energy levels, transition lines, dipole
moments, and transition probabilities. DVR3D uses an exact Hamiltonian, within the Born-Oppenheimer
approximation, and requires potential energy and dipole surfaces to be supplied as input. In
general, it is these which largely determine the accuracy of the resulting calculations
\citep{PolyanskyCSZBe2003}.

The vibrational stage of the DVR3D suite requires an accurate model for the variation of electronic
potential as a function of nuclear geometry. Here we use the \htp\ global potential surface of
\citet{PolyanskyPKT2000}, which used the ultra-high accurate \abin\ data of \citet{CencekRJK1998}
supplemented by extra data points; the surface was constrained at high energy by the data of
\citet{SchinkeDL1980}. We added the \htdp\ adiabatic correction term fitted by
\citet{PolyanskyT1999} to this surface. We also employed Polyansky \&\ Tennyson's vibrational mass
scaling to allow for non-adiabatic corrections to the Born-Oppenheimer approximation: we used
$\mu_\textrm{H}=1.0075372$~u and $\mu_\textrm{D}=2.0138140$~u for the vibrational atomic masses,
and $\mu_\textrm{H}=1.00727647$~u and $\mu_\textrm{D}=2.01355320$~u for the rotational atomic
masses. The accuracy of this model is assessed below.

Calculations were performed in Jacobi coordinates ($r_1, r_2, \theta$), where $r_1$ represents the
diatom distance (H-H), $r_2$ is the separation of the D atom from the center of mass of the diatom,
and $\theta$ is the angle between $r_1$ and $r_2$.  Radial basis functions of Morse oscillator type
were used to model $r_1$ \citep{TennysonS1982}, while spherical oscillators were used to model
$r_2$ \citep{TennysonS1983}. Following \citet{PolyanskyT1999}, the Morse parameters for $r_1$ were
set to $r_e = 1.71$, $D_e = 0.10$ and $\omega_e = 0.0108$, with 20 Gauss-Laguerre grid points.
Parameters of the spherical oscillator functions were set to $\alpha = 0.0$ and $\omega_e = 0.0075$
with 44 Gauss-Laguerre grid points, following extensive tests on convergence of the vibrational
band origins. 36 Gauss-Legendre grid points were used to represent the angular motion. The final
vibrational Hamiltonian matrix used was of dimension 2000.

In the rotation stage, the size of the Hamiltonian matrix, which is a function of $J$, was set to
$1800(J+1)$ following tests on $J=3$ and $J=15$. These tests demonstrated that choosing
sufficiently large values for the rotational Hamiltonian, although computationally expensive, is
crucial for obtaining reliable results. Our aim was to obtain convergence to within 0.01~\wn\ for
all rotation-vibration levels considered. Our tests suggest we achieved this except, possibly, for
some of the highest lying levels. For these levels our basic model, and in particular our
corrections to the Born-Oppenheimer approximation, are not reliable to this accuracy.

To compute the intensity of the vibration-rotation transitions, a dipole moment surface is
required. We used the \abin\ dipole surface of \citet{RohseKJK1994} to calculate the components of
the \htdp\ dipole. In the DIPOLE3 module of DVR3D we set the number of Gauss-Legendre quadrature
points used for evaluating the wavefunctions and dipole surface to 50; this choice is consistent
with the requirement that this parameter should be slightly larger than the number of DVR points
used to calculate the underlying wavefunctions.

The final ST1 line list consists of two main files: one for the levels and the other for the
transitions. These two files are constructed and formatted according to the method and style of the
BT2 water line list \citep{BarberTHT2006}. The total amount of CPU time spent in producing the ST1
list including preparation, convergence tests, and verifying the final results is about 8000 hours.
We used the serial version of the DVR3D suite on PC platforms running Linux (Red Hat) operating
system. Both 32- and 64-bit machines were used in this work although the final data were produced
mainly on 64-bit machines due to the large memory requirement for the high-$J$ calculations.

This variational nuclear motion procedure used for the calculations provides rigorous quantum
numbers: $J$, ortho/para and parity $p$, but not the standard approximate quantum numbers in normal
mode, rigid rotor notation.  We hand labelled those levels for which such quantum numbers could be
assigned in a fairly straightforward fashion.  5000 of these levels are fully designated with 3
rotational ($J,K_a,K_c$) and 3 vibrational ($v_1,v_2,v_3$) quantum numbers, while 341 levels are
identified with rotational quantum numbers only. Some of these assignments are made as initial
guess and hence should be treated with caution. Tables \ref{sampleLevTable} and
\ref{sampleTraTable} present samples of the ST1 levels and transitions files respectively.

\begin{table}
\centering %
\caption{Sample of the ST1 levels file. The columns from left to right are for: index of level in
file, $J$, symmetry, index of level in block, frequency in \wn, $v_1$, $v_2$, $v_3$, $J$, $K_a$,
$K_c$. We used -2 to mark unassigned quantum numbers.}
\label{sampleLevTable} %
{\scriptsize 
\begin{tabular}{rrrrrrrrrrr}
\hline
730 & 1 & 4 &  99 &   18265.61525 & -2 & -2 & -2 & 1 & 1 & 0 \\
731 & 1 & 4 & 100 &   18379.99989 & -2 & -2 & -2 & 1 & 1 & 0 \\
732 & 1 & 4 & 101 &   18499.05736 & -2 & -2 & -2 & 1 & 1 & 0 \\
733 & 2 & 1 &   1 &     131.63473 &  0 &  0 &  0 & 2 & 0 & 2 \\
734 & 2 & 1 &   2 &     223.86306 &  0 &  0 &  0 & 2 & 2 & 0 \\
735 & 2 & 1 &   3 &    2318.35091 &  0 &  1 &  0 & 2 & 0 & 2 \\
736 & 2 & 1 &   4 &    2427.09231 &  0 &  1 &  0 & 2 & 2 & 0 \\
737 & 2 & 1 &   5 &    2490.93374 &  0 &  0 &  1 & 2 & 1 & 2 \\
738 & 2 & 1 &   6 &    3123.27957 &  1 &  0 &  0 & 2 & 0 & 2 \\
739 & 2 & 1 &   7 &    3209.80678 &  1 &  0 &  0 & 2 & 2 & 0 \\
\hline
\end{tabular}
}
\end{table}

\begin{table}
\centering %
\caption{Sample of the ST1 transitions file. The first two columns are for the indices of the two
levels in the levels file, while the third column is for the $\EAC$ coefficients in s$^{-1}$.}
\label{sampleTraTable} %
{\normalsize
\begin{tabular}{rrr}
\hline
     30589 &      29553 &   7.99E-04 \\
     19648 &      18049 &   8.37E-03 \\
      8943 &       7423 &   5.55E-01 \\
      8490 &       7981 &   2.18E-03 \\
     20620 &      22169 &   6.91E-04 \\
     17613 &      15937 &   5.62E-03 \\
     13046 &      11400 &   1.15E-00 \\
     20639 &      20054 &   7.26E-03 \\
     14433 &      17117 &   2.49E-03 \\
     25960 &      28074 &   1.92E-03 \\
\hline
\end{tabular}
}
\end{table}

\section{Results and Validation}

\subsection{Comparison to Experimental and Theoretical Data}

We made a number of comparisons between the ST1 line list and laboratory data found in the
literature. This enabled us to validate our results. The main sources of laboratory measurements
that we used in our comparisons are presented in Table~\ref{ComparisonTable}. The table also gives
statistical information about the discrepancy between our calculated line frequencies and their
experimental counterparts. Table \ref{asvanyTable} presents a rather detailed account of this
comparison for a sample data extracted from one of these data sets, specifically that of
\cite{AsvanyHMKSTS2007}. This table also contains a comparison of relative Einstein \EBC\
coefficients between theoretical values from ST1 and measured values from Asvany \etal. The
theoretical values in this table are obtained from the calculated \EAC\ coefficients using the
relation
\begin{equation}\label{ebcEq}
    \EBC_{lu}=\frac{(2J'+1)\C^{3}\EAC_{ul}}{(2J''+1)_{l}8\pi \h\F^{3}}
\end{equation}
where $\EAC_{ul}$ and $\EBC_{lu}$ are the Einstein A and B coefficients respectively for transition
between upper ($u$) and lower ($l$) states, $J'$ and $J''$ are the rotational quantum numbers for
upper and lower states, $\h$ is Planck's constant, and $\F$ is the transition frequency.

As seen, the ST1 values agree with the measured coefficients of Asvany \etal\ to within
experimental error in all cases. Other comparisons to previous theoretical data, such as that of
\cite{PolyanskyT1999}, were also made and the outcome was satisfactory in all cases.

\begin{table}
\centering %
\caption{Main experimental data sources used to validate the ST1 list. Columns 2 and 3 give the
number of data points and the frequency range of the experimental data respectively, while the last
three columns represent the minimum, maximum, and standard deviation of discrepancies (i.e.
observed minus calculated) in \wn\ between the ST1 and the experimental data sets.}
\label{ComparisonTable} %
{\tiny
\begin{tabular}{lcrrrc}
\hline
    Source &        $N$ & Range (\wn) &       Min. &       Max. &         $\sigma$ \\
\hline
\cite{ShyFW1981}&          9 & 1837 -- 1953 &     $-$0.014 &      0.116 &      0.052 \\
\cite{AmanoW1984}&         27 & 2839 -- 3179 &     $-$0.315 &      0.054 &      0.065 \\
\cite{Amano1985}&         37 & 2839 -- 3209 &     $-$0.024 &      0.054 &      0.019 \\
\cite{FosterMPWPe1986}&   73 & 1837 -- 2603 &     $-$0.134 &      0.213 &      0.067 \\
\cite{FarnikDKPTN2002}&   8 & 4271 -- 4539 &      0.046 &      0.172 &      0.050 \\
\cite{AsvanyHMKSTS2007}& 25 & 2946 -- 7106 &      0.008 &      0.242 &      0.088 \\
\hline
\end{tabular}
}
\end{table}


\begin{table}
\centering %
\caption{Comparison between measured \citep{AsvanyHMKSTS2007} and calculated (ST1) frequencies and
relative Einstein $\EBC$ coefficients for a number of transition lines of \htdp. These coefficients
are normalized to the last line in the table. The absolute B coefficients, as obtained from the A
coefficients of ST1 list using Equation \ref{ebcEq}, are also shown in column 5 as multiples of
$10^{14}$ and in units of str.s.kg$^{-1}$.}
\label{asvanyTable} %
{\scriptsize 
\begin{tabular}{cc|rrc|lr}
\hline
\multicolumn{ 2}{c}{{\bf Transition}} & \multicolumn{ 2}{|c}{{\bf Freq. (\wn)}} & {\bf $\EBC$} & \multicolumn{ 2}{|c}{{Relative \bf $\EBC$}} \\

      Vib. &       Rot. &        Obs. &        ST1 &     ST1   &        Obs. &        ST1 \\
\hline
   (0,3,0) & \rotn 110 $\leftarrow$ \rotn 111 &   6303.784 &   6303.676 &       8.05 &       0.29 &       0.29 \\
   (0,3,0) & \rotn 101 $\leftarrow$ \rotn 000 &   6330.973 &   6330.850 &       8.59 & 0.32$\pm$0.02 &       0.31 \\
   (0,2,1) & \rotn 000 $\leftarrow$ \rotn 111 &   6340.688 &   6340.456 &       7.36 & 0.27$\pm$0.03 &       0.27 \\
   (0,2,1) & \rotn 202 $\leftarrow$ \rotn 111 &   6459.036 &   6458.794 &       9.17 & 0.35$\pm$0.04 &       0.34 \\
   (0,2,1) & \rotn 111 $\leftarrow$ \rotn 000 &   6466.532 &   6466.300 &       27.3 &          1.00 &       1.00 \\
\hline
\end{tabular}
}
\end{table}

\subsection{Partition Function}

The partition function of a system consisting of an ensemble of particles in thermodynamic
equilibrium is given by
\begin{equation}\label{partEq}
    \pf(\T) = \sum_{i}(2J_i+1)\nsdf_{i}e^{(\frac{-\E_{i}}{\BOLTZ\T})}
\end{equation}
where $i$ is an index running over all energy states of the ensemble, $\E_i$ is the energy of state
$i$ above the ground level which has rotational angular momentum $J_i$, $k$ is Boltzmann's
constant, $T$ is temperature, and $g_i$ is the nuclear spin degeneracy factor which is 1 for para
states and 3 for ortho ones.

\begin{figure}
\centering
\includegraphics[scale=0.35]{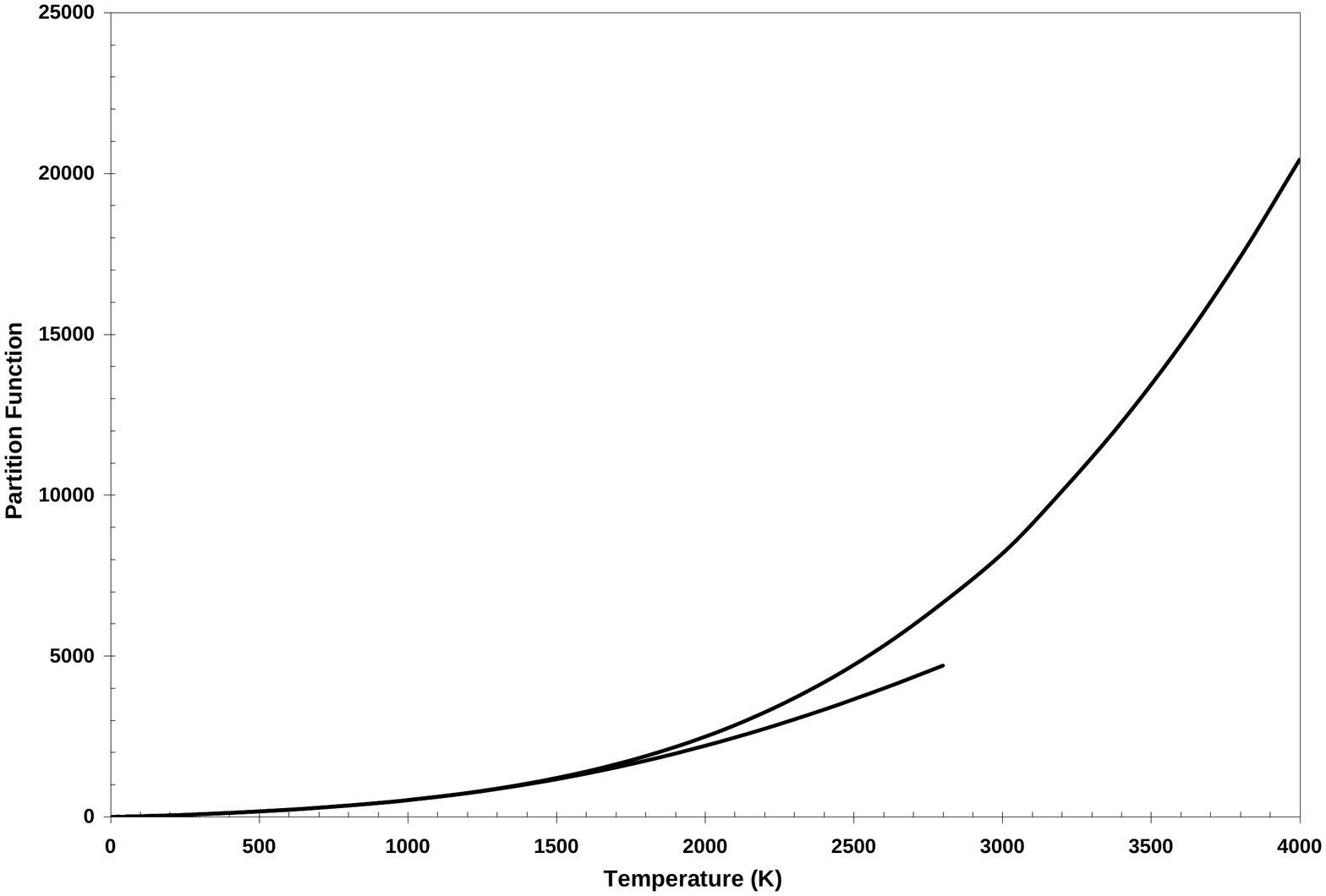}
 \caption
 {The \htdp\ partition functions of ST1 (upper curve) and \citet{SidhuMT1992} (lower curve).
}
 \label{Partition}
\end{figure}

Using the ST1 energy levels we calculated the partition functions of \htdp\ for a range of
temperatures and compared the results to those obtained by \cite{SidhuMT1992}. Table
\ref{sidhuTable} presents these results for a temperature range of 5 -- 4000~K. The results are
also graphically presented in Figure \ref{Partition}. The table and figure reveal that although our
results and those of of Sidhu \etal\ agree very well at temperatures below 1200~K, they differ
significantly at high temperatures and the deviation increases as the temperature rises. This can
be explained by the fact that ST1 contains more energy levels which contribute increasingly at high
temperatures. We therefore expect our partition function to be the more reliable one for $T >
1200$~K.

Using a Levenberg-Marquardt nonlinear curve-fitting routine, we fitted the ST1 partition function
curve to a fifth order polynomial
\begin{equation}\label{partfit}
    \pf(\T) = \sum_{i=0}^5 a_i \T^i
\end{equation}
in temperature and obtained the following coefficients for $\T$ in K:
\begin{eqnarray}
\nonumber  a_0 &=& -0.300315 \\
\nonumber  a_1 &=& +0.094722 \\
\nonumber  a_2 &=& +0.000571 \\
\nonumber  a_3 &=& -3.24415 \times 10^{-7} \\
\nonumber  a_4 &=& +2.01240 \times 10^{-10} \\
           a_5 &=& -1.94176 \times 10^{-14}
\end{eqnarray}
This polynomial is not shown on Figure \ref{Partition} because it is virtually identical to the ST1
curve.

\begin{table}
\centering %
\caption{Partition functions of \htdp\ for a number of temperatures as obtained from
\citet{SidhuMT1992} and ST1 line lists.}
\label{sidhuTable} %
{\normalsize
\begin{tabular}{|r|rr|r|rr|}
\hline
  $\T$(K) & \multicolumn{ 2}{|r|}{Partition Function} &   $\T$(K) & \multicolumn{ 2}{|r|}{Partition Function} \\

           &      Sidhu &        ST1 &            &      Sidhu &        ST1 \\
\hline
         5 &       1.00 &       1.00 &        800 &     347.58 &     347.53 \\
        10 &       1.01 &       1.01 &        900 &     426.24 &     426.24 \\
        20 &       1.07 &       1.28 &       1000 &     515.43 &     516.31 \\
        30 &       2.15 &       2.15 &       1200 &     731.43 &     737.61 \\
        40 &       3.46 &       3.46 &       1400 &    1004.25 &    1026.84 \\
        50 &       5.05 &       5.05 &       1600 &    1339.43 &    1401.52 \\
        60 &       6.82 &       6.82 &       1800 &    1738.89 &    1881.33 \\
        70 &       8.73 &       8.73 &       2000 &    2203.94 &    2487.98 \\
        80 &      10.76 &      10.76 &       2200 &    2735.31 &    3244.83 \\
        90 &      12.90 &      12.89 &       2400 &    3327.58 &    4176.19 \\
       100 &      15.12 &      15.12 &       2600 &    3983.83 &    5306.36 \\
       150 &      27.56 &      27.55 &       2800 &    4698.51 &    6658.69 \\
       200 &      42.03 &      42.02 &       3000 &            &    8254.62 \\
       300 &      76.49 &      76.46 &       3200 &            &   10112.90 \\
       400 &     117.48 &     117.44 &       3400 &            &   12249.00 \\
       500 &     164.51 &     164.53 &       3600 &            &   14674.90 \\
       600 &     218.11 &     217.98 &       3800 &            &   17398.90 \\
       700 &     278.66 &     278.58 &       4000 &            &   20425.60 \\
\hline
\end{tabular}
}
\end{table}

\subsection{Cooling Function}

\begin{figure}
\centering
\includegraphics[scale=0.45]{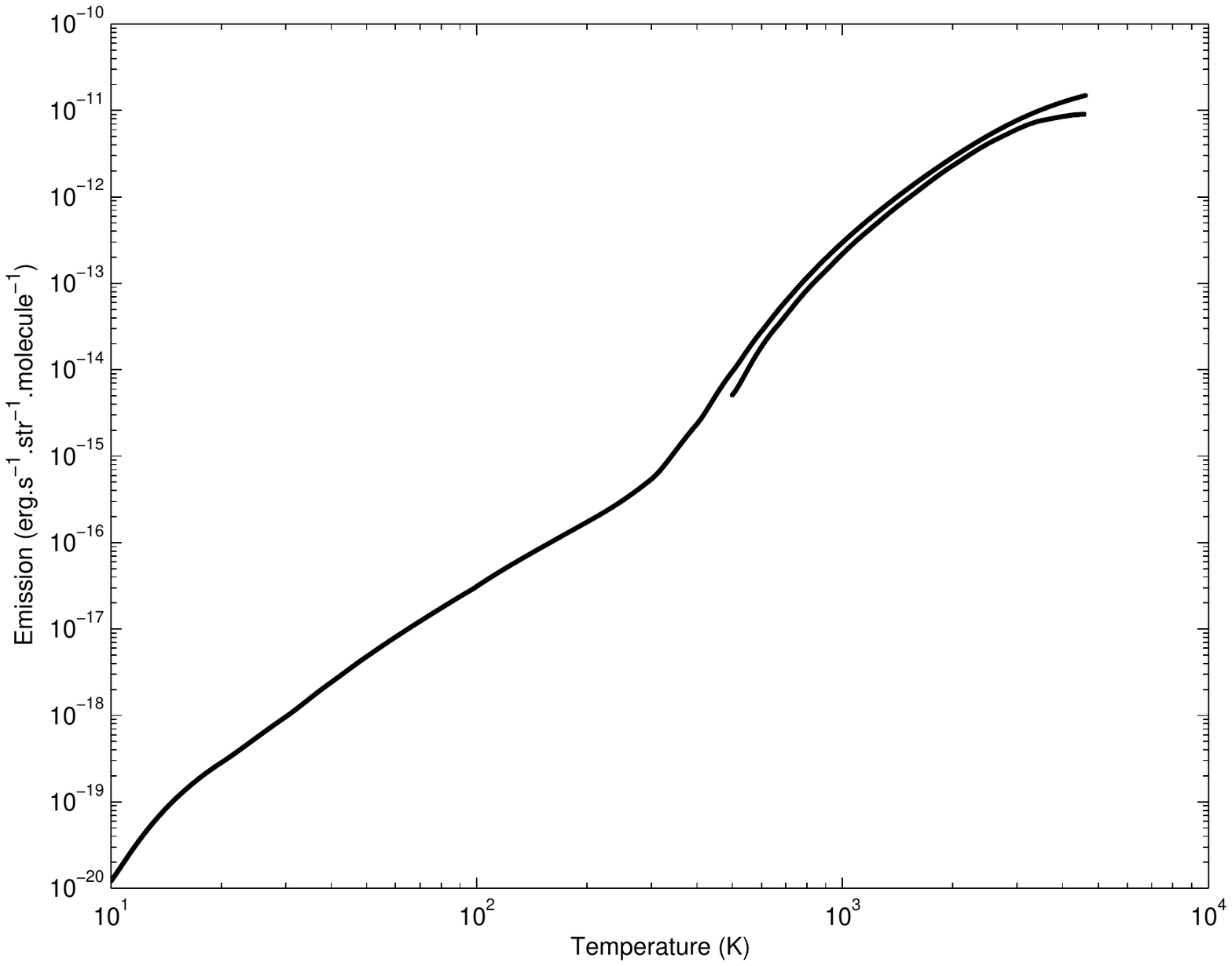}      
 \caption
 {A graph of per molecule emission of \htdp\ (upper curve) and \htp\ (lower curve) as a function of temperature on a
 log-log scale. The \htdp\ curve is obtained from ST1 while the \htp\ curve is obtained from a digitized image from \citet{NealeMT1996}.}
 \label{Cooling}
\end{figure}

ST1 was also used to compute the cooling function of \htdp\ as a function of temperature. The
cooling function $W$ which quantifies the rate of cooling per molecule is given by
\begin{equation} \label{coolEq}
    \cf(\T) = \frac{1}{\pf} \left( \sum_{u,l} \EAC_{ul} (\E_{u}-\E_{l}) (2J_{u}+1)\nsdf_{u}e^{(\frac{-\E_{u}}{\BOLTZ\T})} \right)
\end{equation}
where $u$ and $l$ stand for the upper and lower levels respectively, and $\pf$ is the partition
function as given by Eq.~\ref{partEq}. Figure \ref{Cooling} graphically presents our cooling
function as a function of temperature alongside the \htp\ cooling function of \cite{NealeMT1996}.
As seen, the two curves are close for $T > 600$~K. However the \htdp\ cooling function continues to
be significant at lower temperatures, whereas at these temperatures the cooling curve of \htp\ was
not given by \cite{NealeMT1996} because they considered it too small to be of significance.  This
is of course why the cooling properties of \htdp\ could be important in the early universe.

Using a Levenberg-Marquardt curve-fitting routine, we fitted our \htdp\ cooling function curve to a
fourth order polynomial in temperature over the range 10 -- 4000~K and obtained
\begin{equation} \label{coolfit}
    \cf(\T) =  \sum_{i=0}^4 b_i \T^i
\end{equation}
\begin{eqnarray}
\nonumber  b_0 &=& -1.34302 \times 10^{-16} \\
\nonumber  b_1 &=& +1.56314 \times 10^{-17} \\
\nonumber  b_2 &=& -2.69764 \times 10^{-19} \\
\nonumber  b_3 &=& +7.03602 \times 10^{-22} \\
           b_4 &=& -1.10821 \times 10^{-25}
\end{eqnarray}
The fit was essentially perfect on a linear scale.

\subsection{Synthetic Spectra}

\begin{figure*}
\centering
\includegraphics[width=1.0\textwidth]{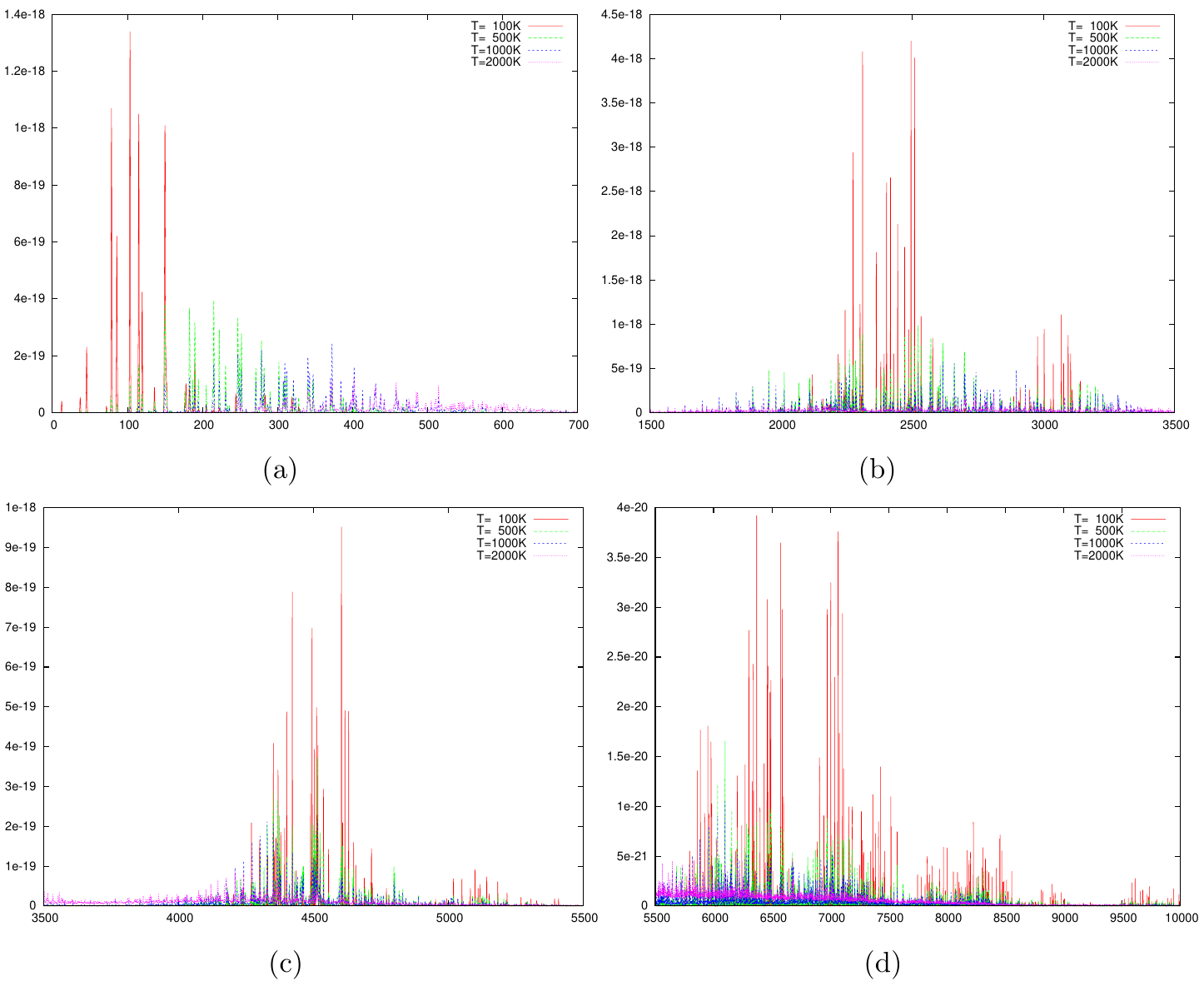}      
 \caption
 {A graph of integrated absorption coefficient in cm.molecule$^{-1}$ (on $y$ axis) as a function of wave number
 in \wn\ (on $x$ axis) within the range $0-10000$~\wn\ for $\T=$ 100, 500, 1000 and 2000~K.}
 \label{SynthSpec}
\end{figure*}

One of the main uses of line lists such as ST1 is to generate temperature-dependent synthetic
spectra. We generated spectra from ST1 using the Spectra-BT2 code, which is described by
\cite{BarberTHT2006} and is a modified version of the original spectra code of \cite{Tennyson1993}.
Figure~\ref{SynthSpec} (a) shows a temperature-dependent \htdp\ absorbtion spectrum for the region
largely associated with pure rotational transitions, while Figures~\ref{SynthSpec} (b), (c) and (d)
show the corresponding spectrum for the vibrational region. As is usual with rotation-vibration
spectra, \htdp\ spectra show a strong dependence on temperature.

\section{Conclusions}

In this study we present a new line list (ST1) for the triatomic ion \htdp. The list, which can be
obtained from the Centre de Donn\'{e}es astronomiques de Strasbourg (CDS) database at
ftp://cdsarc.u-strasbg.fr/pub/cats/VI/130, comprises over 33 thousand rotational-vibrational energy
levels and more than 22 million transition lines archived within two files. The tests performed on
the line list suggest that, although it is based entirely on \abin\ quantum mechanics, it should be
accurate enough for almost all astronomical purposes. The one possible exception is for predicting
the frequency of pure rotational transitions which are often needed to high accuracy and which are
therefore better obtained from measured frequencies using combination differences.

\section{Acknowledgement} \label{Acknowledgements}

We thank Lorenzo Lodi for his help with the DVR3D code, and Steven Miller for encouraging to
calculate \htdp\ line list and for helpful discussions.

\label{lastpage}

\end{document}